\documentclass{article}

\usepackage{fix-cm}

\usepackage[margin=1.5in]{geometry}
\usepackage[utf8]{inputenc} %
\usepackage{textcomp}
\usepackage[american]{babel} %
\usepackage{color} %
\usepackage{amsmath} 
\usepackage{amsfonts}
\usepackage[authoryear]{natbib} %
\usepackage{graphicx} 
\usepackage{bm}
\usepackage{booktabs}
\usepackage[affil-it]{authblk} 
\usepackage{lmodern} %
\usepackage{setspace}
\newcommand{\argmin}{\mathop{\textrm{argmin}}}

\begin{document}

\date{}

\title{Non-parametric estimation of the first-order Sobol indices with
	bootstrap bandwidth}

\author{Maikol Solís} %
\affil{Centro de Investigación en Matemática Pura y
	Aplicada, Escuela de Matemática. \\
	Universidad de Costa Rica, Costa Rica.\\
	Email: maikol.solis@ucr.ac.cr}

\maketitle

\begin{abstract}

  Suppose that \(Y = \psi(X_1, \ldots, X_p)\), where \((X_1,\ldots, X_p)^\top\) are
  random inputs, \(Y\) is the output, and \(\psi(\cdot)\) is an unknown link
  function. The Sobol indices gauge the sensitivity of each \(X\) against \(Y\)
  by estimating the regression curve's variability between them. In this paper,
  we estimate these curves with a kernel-based method. The method allows to
  estimate the first order indices when the link between the independent and
  dependent variables is unknown. The kernel-based methods need a bandwidth to
  average the observations. For finite samples, the cross-validation method is
  famous to decide this bandwidth. However, it produces a structural bias. To
  remedy this, we propose a bootstrap procedure which reconstruct the model
  residuals and re-estimate the non-parametric regression curve. With the new
  set of curves, the procedure corrects the bias in the Sobol index. To test
  the developed method, we implemented simulated numerical examples with
  complex functions.

	\textbf{Keywords:} Sobol indices, Sensitivity Analysis, Non-parametric
	estimator, Finite-sample bias, Bootstrap bandwidth.

	\textbf{MSC2010:} 62G08, 62F40, 93B35.

\end{abstract}

\section{Introduction}

Researchers, technicians or policy-makers often support their decisions on
complex models. They have to process, analyze and interpret them with the data
available. In normal conditions, those models include many variables and
interactions. One choice to overcome these issues is selecting the most
relevant variables of the system. In this way, we will gain insight on the
model, and we will discover the main characteristics. Still, we have to produce
a stable approximation of the model to avoid large variations on the input
given by small perturbations on the output. The analyst, however, has to
confirm, check and improve the model.

The typical situation assumes a set of inputs variables
\(\bm{X}=(X_1, \ldots, X_p)^\top\in\mathbb{R}^p\) producing an output \(Y\in\mathbb{R}\)
related by the model
\begin{equation}
	\label{eq:sobol:model-sobol}
	Y = \psi(X_1, \ldots, X_p).
\end{equation}

The function \(\psi(\cdot)\) could be unknown and complex. Sometimes, a
computer code can gauge it (e.g., \citet{oakley2004probabilistic}). Also, we
can replace the original model by a low fidelity approximation called a
\emph{meta-model} (see \citet{box1987empirical}). The problems related to this
formulation extend to engineering, biology, oceanography and others.

Given the set of inputs \((X_1, \ldots, X_p)\) in the model defined in
model~\eqref{eq:sobol:model-sobol}, we can rank them according different
criteria. Some examples are: the screening method
(\cite{cullen1999probabilistic, campolongo2011screening}), the
automatic differentiation (\citet{rall1980applications,
	carmichael1997sensitivity}), the regression analysis
(\citet{draper1981applied, devore1996statistics}) or the response
surface method (\citet{montgomery1995response, goos2002optimal}).

Inspired by an ANOVA (or Hoeffding) decomposition,
\citet{sobol1993sensitivity}, split down the variance of the model in
partial variances. They are generated by the conditional expectations
of \(Y\) giving each input \(X_i\) for \(i=1, \ldots, p\). The partial
variances represent the uncertainty created by each input or its
interactions. Dividing each partial variance by the model total
variance, we get a normalized index of importance. We call the
first-order Sobol indices to the quantities,
\begin{equation*}
	S_i = \frac{\mathrm{Var}(\mathbb{E}[Y\vert X_i])}{\mathrm{Var}(Y)}
	\quad \text{for} \quad i=1, \ldots, p.
\end{equation*}

Notice that \(\mathbb{E}[Y\vert X_i]\) is the best approximation of \(Y\) given
\(X_i\). Thus, if the variance of \(\mathbb{E}[Y\vert X_i]\) is large, it means a
large influence of \(X_i\) into \(Y\).

The Sobol indices determine the most relevant and sensible inputs on the
model. We can establish indices that measure the interactions between
variables or the total effect of a certain input in the whole model. We
refer the reader to \citet{saltelli2000sensitivity} for the exact
computation of higher-order Sobol indices.

The main task with the Sobol indices relay in its computation.
Monte-Carlo or quasi Monte-Carlo methods propose sampling the model (of
the order of hundreds or thousands) to get an approximation of its
behavior. For instance, the Fourier Amplitude Sensitivity Test (FAST) or
the Sobol Pick-Freeze (SPF) \citet{cukier1973study,cukier1978nonlinear}
created the FAST method which transforms the partial variances in
Fourier expansions. This method allows the aggregated and simple
estimation of Sobol indices in an escalated way. The SPF scheme
regresses the model output against a pick-frozen replication. The
principle is to create a replication holding the interest variable
(frozen) and re-sampling the other variables (picked). We refer to
reader to \citet{sobol1993sensitivity}, \citet{sobol2001global} and
\citet{janon2013asymptotic} Other methods include to
\citet{Ishigami1990} which improved the classic Monte-Carlo procedure by
resampling the inputs and reducing the whole process to only one
Monte-Carlo draw. The paper of \citet{Saltelli2002} proposed an
algorithm to estimate higher-order indices with the minimal computation
effort.

The Monte-Carlo methods suffer from the high-compu\-tational stress in
its implementation. For example, the FAST method requires estimate a set
of suitable transformation functions and integer angular frequencies for
each variable. The SPF scheme creates a new copy of the variable in each
iteration. For complex and high-dimensional models, those techniques
will be expensive in computational time.

One limitation of the methods mentioned before is a complete
identification of the link function \(\psi(\cdot)\) between the inputs and the
output. It means, the analyst has to have the exact link function or an
alternative algorithm which produce the outcome. Otherwise, if we
have only available a data set with explanatory and response variables
the question remains on finding the most influential explanatory variables without any additional information.

This article proposes an alternative way to com\-pu\-te the Sobol indices.  In
particular, we will take the ideas of \citet{zhu1996asymptotics} and we shall
apply a non-parametric Nadaraya-Watson to estimate the value \(S_i\) for \(i=1,
\ldots, p\). With this estimator, we avoid the stochastic techniques, and we
use the data to fit the non-parametric model. If the joint distribution of
\((X_i, Y)\) is twice differentiable, the non-parametric estimator of \(S_i\),
has a parametric rate of convergence.  Otherwise, we will get a non-parametric
rate of convergence depending on the regularity of the density. The classic way
to estimate the bandwidth for the non-parametric estimator is through
cross-validation. We will implement a bootstrap procedure to remove the
structural bias generated by cross-validation bandwidth.  

The article follows this framework: We start with some preliminaries in
Section~\ref{sec:preliminaries}. In Section~\ref{sec:sobol:methodology} we will
propose the non-parametric estimator for the first-order Sobol indices.  The
method to calibrate bootstrap bandwidth selection in
Section~\ref{sec:choice-bandw-sobol}. We show our method with two numerical
examples in Section~\ref{sec:results}. Finally, Section~\ref{sec:conclusions},
we will expose the conclusions and discussion.

\section{Preliminaries}\label{sec:preliminaries}

The sensitivity analysis is \textit{``the
	study of how uncertainty in the output of a model can be apportioned to
	different sources of uncertainty in the model input''}
(\citet{Saltelli2009}). In a modeling environment, include the step to
validate if all the variables explain something relevant to the model is
crucial. A complete analysis setting allows to review, to validate and
to simplify any model.

A popular method to identify those variables is the Sobol indices
method. The method proposed by \citet{sobol1993sensitivity} using an
orthogonal decomposition of functions in the unitary cube. The result
separates the regression effects and then estimates how much variability
contributes to explain a model.

Formally, if \(y = \psi(x_{1}, \ldots, x_{p})\) is a squared integrable
function with domain \(\mathcal{D} = {[0,1]}^{p}\), then
\begin{multline}
	\label{eq:sobol-descomposition}
	\psi(x_{1}, \ldots, x_{p}) = \psi_{0} + \sum_{i} \psi_{i}(x_{i}) + \sum_{ij}
	\psi_{ij}(x_{i}, x_{j}) + \\
	\sum_{ijk} \psi_{ij}(x_{i}, x_{j}, x_{k}) + \cdots+ \psi_{12\cdots p}(x_{1},
	\ldots, x_{p})
\end{multline}
where the term \(\psi_0\), is constant and the functions \(\psi_{i}\),
\(\psi_{ij}\) and so on are also square integrable over its respective
domain. This decomposition has \(2^{p}\) terms.

\citeauthor{sobol1993sensitivity} showed the
expression~\eqref{eq:sobol-descomposition} has a unique representation
when each component are centered and pairwise orthogonal.

Setting \((X_{1}, \ldots, X_{p})\) and \(Y = \psi(X_{1}, \ldots, X_{p})\),
Equation~\eqref{eq:sobol-descomposition} are the split contributions of
the inputs to the output \(Y\) due to the interactions of: none variables,
one variable, two variables and so on. Note that if we take the
conditional expectation to the variable \(Y\), we could reinterpret
Equation~\eqref{eq:sobol-descomposition} as,
\begin{align}
	\label{eq:orthogonal-sobol}
	\begin{split}
		\psi_0 & = \mathbb{E}[Y] \\
		\psi_i(X_{i}) & = \mathbb{E}[Y\vert X_i] - \psi_{0} \\
		\psi_{ij}(X_{i}, X_{j}) & = \mathbb{E}[Y\vert X_i, X_j] \\
		&\qquad - \psi_i(X_{i}) - \psi_j(X_{j}) - \psi_0, 
	\end{split}
\end{align}
and so on for the other functions.

The variance of each term in~\eqref{eq:orthogonal-sobol} measures the
relevance of each set of variables into the model. In this case, the Sobol indices are
the normalized version by the total variance of \(Y\). The first order
effects remain as,
\begin{align}
	\label{eq:sobol-indices}
	S_{i} & = \frac{\mathrm{Var}(\mathbb{E}[Y\vert X_{i}])}
	{\mathrm{Var}(Y)}.
\end{align}

The total contribution of a variable is measured with the quantity,
\begin{equation*}
	S_{T_{i}} = 1 - \frac{\mathrm{Var}(\mathbb{E}[Y\vert X_{\sim i}])}
	{\mathrm{Var}(Y)}
\end{equation*}
where \(X_{\sim i}\) means all the variables except the variable
\(X_{i}\).

In our framework, we are interested in the first order Sobol indices
\(S_{i}\) estimated using a non-parametric method. The method depends
solely on the data available of the inputs and the output. It ignores
the particular form of the link function used to generate the output.
This features will allow us to estimate Sobol indices in models when the
relationship between \(X_i\)'s and \(Y\) are unknown.

\section{Methodology}\label{sec:sobol:methodology}

In our context we suppose that \(\boldsymbol{X}_k = (X_{1k}, \ldots,
X_{pk})^\top\) are independent and identically distributed observations from
the random vector \(\boldsymbol{X} = (X_{1}, \ldots, X_{p})^\top\). Also, \(Y_k
= m(X_{1k}, \ldots, X_{pk})\) for \(k=1,\ldots, n\) where \(m\) is the link
functions is defined in Equation~\eqref{eq:sobol:model-sobol}. We denote
by \(f(x_i, y)\) the joint density of the couple \((X_i, Y)\). Let \(f_i
(x_i) = \int_{\mathbb{R}^p} f(x_i, y)dy\) be the marginal density function of
\(X_i\) for \(i=1,\ldots, p\).

Recall Sobol indices definition presented in the introduction,
\begin{equation}
	\label{eq:sobol:def-sobol}
	S_i = \frac{\mathrm{Var}(\mathbb{E}[Y\vert X_i])}{\mathrm{Var}(Y)}
	= \frac{{\mathbb{E}[\mathbb{E}[Y\vert X_i]]}^{2} - {\mathbb{E}[Y]}^{2}}{\mathrm{Var}(Y)}
	\quad \text{for} \quad i=1, \ldots, p.
\end{equation}

We have expanded the variance of the numerator to simplify the
presentation. Notice we can estimate the terms \(\mathbb{E}[Y]\) and \(\mathrm{Var}(Y)\) in
equation~\eqref{eq:sobol:def-sobol} by their empirical counterparts
\begin{align}
	\label{eq:sobol:def-variance-empiric-Y}
	\overline{Y}     & = \frac{1}{n} \sum_{k=1}^n Y_k \\
	\intertext{and}
	s_Y^2 & = \frac{1}{n-1}
	\sum_{k=1}^n {(Y_k-\overline{Y})}^2.
\end{align}

The term \(\mathbb{E}[{\mathbb{E}[Y\vert X_i]}^2]\) requires more effort to estimate. For
any \(i=1,\ldots, p\) we introduce the following notation,
\begin{align*}
	V_i = \mathbb{E}[{\mathbb{E}[Y\vert X_i]}^2]
	         & = \int {\left(\frac{\int y \,
			f(x_i, y)\, dy}{f_i(x_i)}\right)} ^2 f_i(x_i)\, dx_i \\
	         & = \int {\left(\frac{g_i(x_i)}{f_i(x_i)}\right)}^2
	f_i(x_i)\, dx_i,                                             \\
	\intertext{where}
	g_i(x_i) & = \int y \, f(x_i, y)\, dy.
\end{align*}

We will use a changed version of the non-parametric estimator developed
in \citet{loubes2013rates}. This paper estimates the conditional
expectation covariance for reduce the dimension of a model using the
sliced inverse regression method.

We will estimate the functions \(g_i(x)\) and \(f_i(x)\) by their
non-parametric estimators,
\begin{align}
	\hat{g}_{i, h}(x)
	  & = \frac{1}{nh} \sum_{l=1}^{n} Y_{l}
	K\left(\frac{x - X_{il}}{h}\right), \label{eq:sobol:g-i} \\
	\hat{f}_{i, h}(x)
	  & = \frac{1}{n h} \sum_{l=1}^{n} K\left(\frac{x -
		X_{il}}{h}\right).  \label{eq:sobol:f-X}
\end{align}

The non-parametric estimator for \(V_i\) is,
\begin{equation}
	\label{eq:sobol:def-variance-conditional}
	\widehat{V}_{i}(h)
	= \displaystyle \frac{1}{n} \sum_{k=1}^{n}
	{\left(\frac{\hat{g}_{i, h}(X_{ik})}
	{\hat{f}_{i, h}(X_{ik})}\right)}^2.
\end{equation}

Thus, we can gather the
estimators~\eqref{eq:sobol:def-variance-empiric-Y}
and~\eqref{eq:sobol:def-variance-conditional} and define the
non-parametric estimator for \(S_i\) as,
\begin{equation}
	\label{eq:sobol:def-estimator-sobol-index}
	\widehat{S}_i(h) = \frac{\widehat{V}_i(h)
		- \overline{Y}^2} {s_Y^2}.
\end{equation}

The estimator~\eqref{eq:sobol:def-estimator-sobol-index} provides a
direct way to estimate the first-order Sobol index \(S_i\).

Notices that the estimator \(\widehat{S}_i(h)\) relies on the choice of
an adequate bandwidth \(h\).  The next Section we will propose an
algorithm to select the bandwidth which also minimize the structural
bias caused by the nature of the estimator.

\section{Choice of bandwidths for Sobol indices}%
\label{sec:choice-bandw-sobol}

The last section presented the methodology to estimate the first order
Sobol indices using a non-parametric framework. However, the choice of
the bandwidth \(h\) remains as the crucial step to estimate accurately
\(\widehat{S}_{i}(h)\). The main issue is to estimate the regression curve
\(m_{i}(x)\) define by \(\mathbb{E}[Y\vert X_{ik} = x]\).

We can fit this curve with the data available minimizing the least
squares criteria.
\begin{equation*}
	\mathrm{LS} = \frac{1}{n} \sum_{k=1}^{n}
	{\left\{Y_{k} - m_{i} (X_{ik}) \right\}}^{2}.
\end{equation*}

As before, we estimate \(m_{i} (X_{ik})\) by
\begin{equation*}
	\hat{m}_{i, h}(x) = \frac{\hat{g}_{i, h}(x)}{\hat{f}_{i, h}(x)}
\end{equation*}
where \(\hat{g}\) and \(\hat{f}\) were defined in
Equations~\eqref{eq:sobol:g-i} and~\eqref{eq:sobol:f-X}. But, there
exist a problem because this method uses twice the data to calibrate and
verify the model. The cross-validation method estimate the prediction
error removing one by one the observations and recalculating the model
with the remaining data. The estimator is called \textit{leave-one-out}
estimator with the expression
\begin{equation*}
	\hat{m}_{i, h, -k} (X_{k}) =
	\frac{\sum_{j\neq k} K_{h}(X_{ij}-X_{ik}) Y_{k}}
	{\sum_{j\neq k} K_{h}(X_{ij}-X_{ik})}.
\end{equation*}

Afterwards, we can build a new version of the least squares error
\begin{equation}
	\label{eq:MLS}
	\operatorname{CVLS}(h) = \frac{1}{n} \sum_{k=1}^{n} {\left\{Y_{k} -
	\hat{m}_{i, h, -k} (X_{k}) \right\}}^{2},
\end{equation}
and find the optimal bandwidth
\begin{equation}
  \label{eq:hCV}
  \hat{h}_{CV} = \argmin_{h} \mathrm{CVLS}(h).
\end{equation}

Finally, estimate \(\widehat{S}^{\mathrm{CV}}_{i}(\hat{h}_{CV})\), for the
interested reader, \citet{hardle2004nonparametric} has the detailed procedure.

However, even if the cross-validation is asymptotically unbiased, those
estimators have a relatively large finite-sample bias. The works from
\citet{Faraway1990}, \citet{Romano1988} and \citet{Padgett1986}
established the same behavior studying non-parametric estimators for the
density, quantiles and the mode respectively. This problem arises on the
non-parametric-based models, as it was exemplified by
\citet{Hardle1993}. One solution is remove the bias part of the estimate
by bootstrapping, following the ideas in \citet{Racine2001}.

The procedure starts with the residuals for the variable \(Y\) with
respect to its non-parametric estimate counterpart with some bandwidth
\(h_{0}\),
\begin{equation*}
	\hat{\varepsilon}_{ik} = Y_{k} - \hat{m}_{i, h_{0}}(X_{ik})
	\quad k=1,\ldots, n.
\end{equation*}

For practical purpose $h_0 = \hat{h}_\textrm{CV}$ defined in Equation~\eqref{eq:hCV}. 

Denote the conditional variance of $\hat{\varepsilon}_{ik}$ given the observation
$X_{ik}$, i.e.\! $\mathrm{Var}\left(\hat{\varepsilon}_{ik} \vert X_{ik} \right)$  as
$\sigma_\varepsilon\left(X_{ik} \right)$. %

The  residuals are then normalized by the transformation
\begin{equation*}
	\hat{\nu}_{ik} = \frac{\hat{\varepsilon}_{ik} - \bar{\varepsilon}_i}
	{\sigma_{\varepsilon}(X_{ik})}, \quad k = 1, \ldots, n
\end{equation*}
where \(\bar{\varepsilon}_i\) is the arithmetic mean of
\(\hat{\varepsilon}_{ik}\).
Normalizing the $\hat{\varepsilon}_{ik}$ we produce 
random variables $\hat{\nu}_i$ with mean 0 and variance 1  in each point of the sample.

Denote \(\nu_{i}^{*}\) a bootstrap sample taken from
\({\{\hat{\nu_{i}}\}}_{k=1}^{n}\). The bootstrap sample takes draws with
replacement from the empirical distribution of \(\hat{\nu}_{ik}\).
The technique overcomes the he\-te\-ro\-sce\-das\-ti\-ci\-ty issue by creating multiple
versions of the variable $\nu_{ik}$ and spreading this \textit{pure noise} across all
the sample. For example, \citet{Zhao2017} handles the heteroscedasticity for models with varying coefficients resampling the residuals through a bootstrap technique.

Based on the noise spread over all the sample points,  we
reconstruct the response variable defining 
\begin{equation} Y_{ik}^{*} =
\hat{m}_{i,h_0}(X_{ik}) + \sigma_{\varepsilon}(X_{ik})\nu_{ik}^{*},
\end{equation} 
as a bootstrap sample of \(Y\). Here, we take a base mean
function \(\hat{m}_{ih}(x)\) with the resampled noise \(\nu_{i}^{*}\)
multiplied by $\sigma_\varepsilon\left(X_{ik} \right)$.  Notice that
while the random variable $\nu_{ik}^\star$ distribute the influence of
the noise across all the sample, the value
$\sigma_\varepsilon\left(X_{ik} \right)$ fixes the structural variance
from the original sample in each point. In other words, we are making a
new response variable with the same conditional variance in each point
of the sample, but with a different randomness.  
This new sample of \(Y\) depends on index \(i\), because the errors were
taken from the residuals between \(Y\) and \(\hat{m}_{i, h}(X_{ik})\).
Thus, for \(b = 1,\ldots, B\) we take \((X_{i}, Y_{i}^{(b)}) =
\{(X_{ik}, Y_{ik}^{*}), k=1,\ldots, n\}\) a sample with replacement from
\((X_{i}, Y)\). For each sample \((X_{i}, Y_{i}^{(b)})\), estimate the
regression curve \begin{equation*} \hat{m}^{(b)}_{i, h} (x) =
  \frac{\sum_{j\neq k} K_{h}(X_{ij}-x) Y^{(b)}_{ik}} {\sum_{j\neq k}
K_{h}(X_{ij}-x)} \end{equation*}

The regression is performed conditionally under the design sequence
\(X_{i}\)'s which are not resampled. The curves for each bootstrap
sample depend on a unknown bandwidth $h$.

\begin{figure}[htp]
  \centering \includegraphics[width = \textwidth]{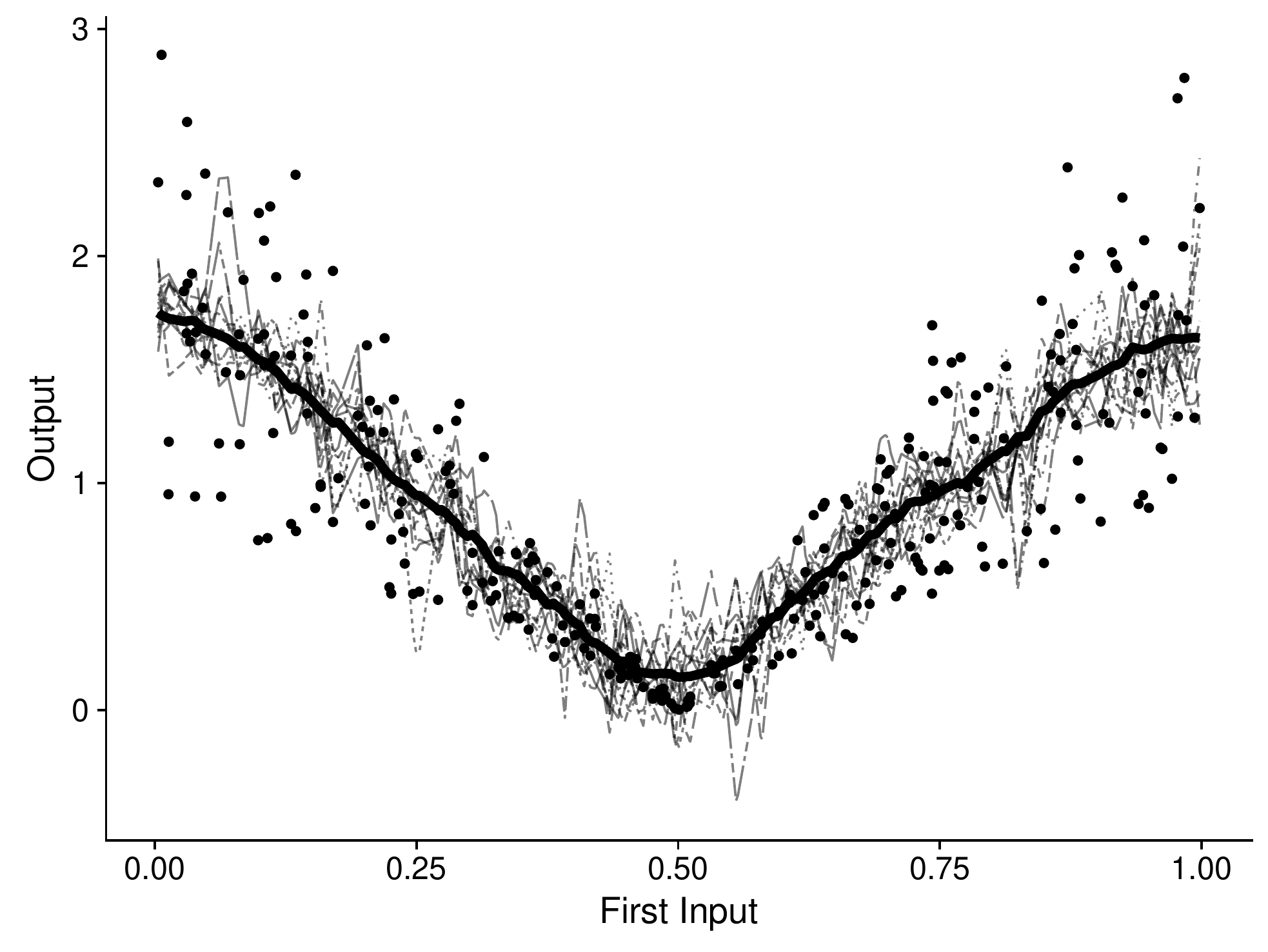}
  \caption{Data points (black dots), bootstrap curves (gray lines) and
  mean curve (black solid line) allows to from \(n = 300\) observations of the
\(g\)-Sobol (Equation~\eqref{eq:g-sobol}) for the first input \(X_{1}\)
against the output \(Y\).}\label{fig:boot-np-mean} 
\end{figure}

As an example, for the \(g\)-Sobol
model explained in Section~\ref{sec:simulations} we took the first input
\(X_{1}\) against the output \(Y\).  Figure~\ref{fig:boot-np-mean}
presents 100 curves generated by the bootstrap procedure and their mean.
Notice that each bootstrap curve presents more variance than the mean
curve. This behavior allows to capture all the irregularities in the
model and produces a better fit of the data.  

Obtaining the \(B\) regression curves, we can now compare the distance
in mean square from the observational output and the bootstrap mean
curve.  Thus, an improvement to the least-square error presented in
Equation~\eqref{eq:MLS} is, \begin{equation*} \mathrm{BLS}_{i}(h)
  =\frac{1}{n}\sum_{k=1}^{n}{\left\{Y_{k} - \frac{1}{B} \sum_{b = 1}^{B}
  \hat{m}^{(b)}_{ih}(X_{ik}) \right \}}^{2}.  \end{equation*} We call it
  \textit{Bootstrap Least-Square} criterion. The second term in the last
  expression produces the mean curve generated from the \(B\) bootstrap
  curves. The function \(\mathrm{BLS}\) reaches its smallest value at,
  \begin{equation*} \hat{h}_{boot} = \argmin_{h} \mathrm{BLS} (h).
  \end{equation*} Getting the bandwidth \(\hat{h}_{boot}\) estimated, it
  only left re-estimate the Sobol index with the new bootstrap
  structure, \begin{equation*}
    \widehat{S}_{i}^{\mathrm{Boot}}(\hat{h}_{boot}) =
    \frac{\mathrm{Var}\left(\displaystyle \frac{1}{B} \sum_{b = 1}^{B}
    \hat{m}^{(b)}_{i,  \hat{h}_{boot}} (X_{k})\right)}
  {\mathrm{Var}(Y)}.  \end{equation*} The procedure captures the
  different irregularities in the data, without having an explicit
  functional form of the model. The procedure summarizes those
  irregularities in a mean curve and create a corrected Sobol index for
  each variable.

\section{Numerical Illustrations}\label{sec:results}
\subsection{Simulation study}\label{sec:simulations} 

Simulations were
performed to determine the quality Sobol index estimator using the
classic cross-validation and bootstrap procedures. In all the
simulations we will take \(n\) equal to \(100\), \(200\), \(300\) for
each case.  We repeated the experiment 100 times selecting different
samples in each iteration. In the bootstrap case, 100 draws were taken
in each iteration. The inputs are uniform random variables for the
chosen configuration. For all simulations, the algorithm executed the
non-parametric regression with second and fourth Epanechnikov kernel.
These kernels are defined by \(K(u) = (3/4) \left(1-u^{2}\right)\) and
\(K(u)= (45/32) \left(1-(7/3) u^{2}\right) \left(1-u^{2}\right)\) for
\(\vert u\vert \leq 1\) in both cases. The purpose of including fourth
order Kernels is to reduce the bias, giving a smoother structure to the
model (for further details, see \citet{Tsybakov2009}). In this sense, we
could compare the fourth order kernel with our procedure.  For a detailed
explanation on higher order kernels see \citet{Hansen2005}.

The software used was \textit{R} (\citet{RCoreTeam2017}), along the
package \texttt{np} (\citet{Hayfield2008}) for all the non-parametric
estimators and the routine \texttt{optimize} to minimize the function
\(\mathrm{BLS}\). The setting considered is called \textit{g}-Sobol and
defined by,

\begin{gather}
	f(x_{1}, \ldots, x_{d}) = \prod_{i=1}^{p} \frac{\vert 4x_{i} -2
		\vert + a_{i}}{1 + a_{i}} \label{eq:g-sobol}\\
	Y = f(X_{1}, \ldots, X_{d}) \text{with} X_{i}\sim\mathrm{Uniform}(0,1)
	\nonumber
\end{gather}
where the \(a_{i}\)'s are positive parameters. The \textit{g}-Sobol is a
strong nonlinear and non monotonic behavior function. As discussed by
\citet{Saltelli2008}, this model has exact first order Sobol indices
\begin{equation*}
	\displaystyle S_{i} =
	\frac{1}{3{(1+a_{i})}^{2}} \bigg/
	\left(-1 + \prod_{k=1}^{p} \left(  1 +
	\frac{1}{3{(1+a_{i})}^{2}}\right) \right).
\end{equation*}
For each \(i\), the lower is the value of \(a_{i}\), the higher is the
relevance of \(X_{i}\) in the model. The parameters used in the
simulations are \(a_{1} = 0\), \(a_{2}=1\), \(a_{3} = 4.5\), \(a_{4}= 9\),
\(a_{5}=a_{6} = a_{7}=a_{8}=99\) with Sobol indices \(S_{1} = 0.7162\),
\(S_{2} = 0.1790\), \(S_{3}= 0.0237\), \(S_{4} = 0.0072\) and \(S_{5}, = S_{6}
= S_{7} = S_{8} = 0.0001\).

To compare our method, we estimate in parallel the following methods for
Sobol indices: B-spline smoothing (\citet{Ratto2010}), and the schemes
by Sobol (\citet{sobol1993sensitivity}), Saltelli
(\citet{Saltelli2002}), Mauntz-Kuncherenko (\citet{Sobol2007}),
Jansen-Sobol (\citet{Jansen1999}), Martinez and Touati
(\citet{Baudin2016}, \citet{Touati2016}), Janon-Monod
(\citet{Makowski2006}), Mara (\citet{AlexMara2008}) and Owen
(\citet{Owen2013}). Those methods do not represent an exhaustive list,
but give wide point of comparison between estimators. All methods
estimated---except the B-splines---need the prior knowledge of the link
function in equation~\eqref{eq:g-sobol} between the input \(X\) and the
output \(Y\).

\begin{figure*}[htp]
	\centering \includegraphics[width = \textwidth]{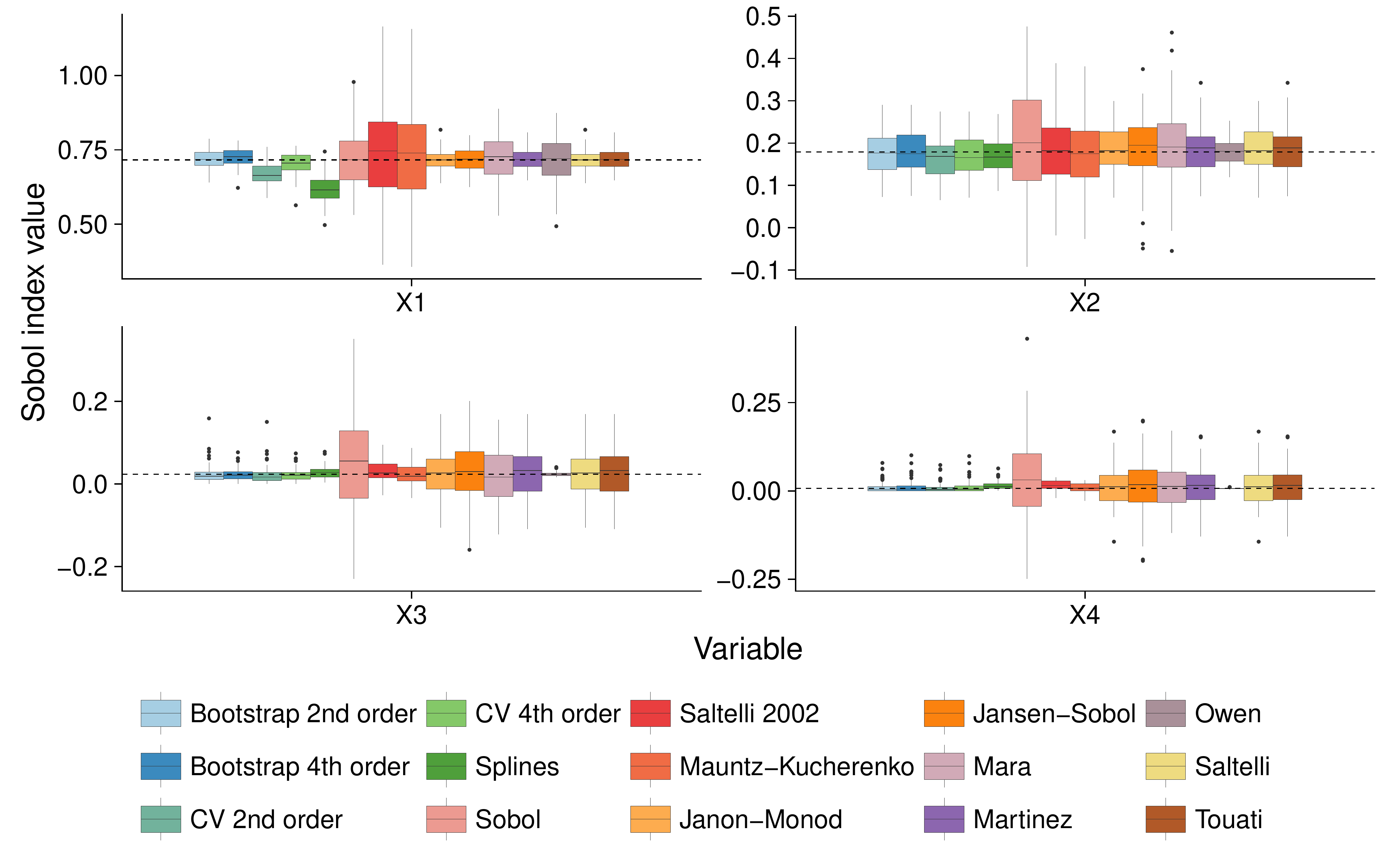}
	\caption{Estimated values from the 100 iterations for the
		\textit{g}-Sobol model across different methodologies. The horizontal
		dashed lines represent the theoretical values for each Sobol index.}%
	\label{fig:all-sobol}
\end{figure*}

\begin{figure*}[htp]~\label{fig:np-sobol}
	\centering \includegraphics[width = \textwidth]{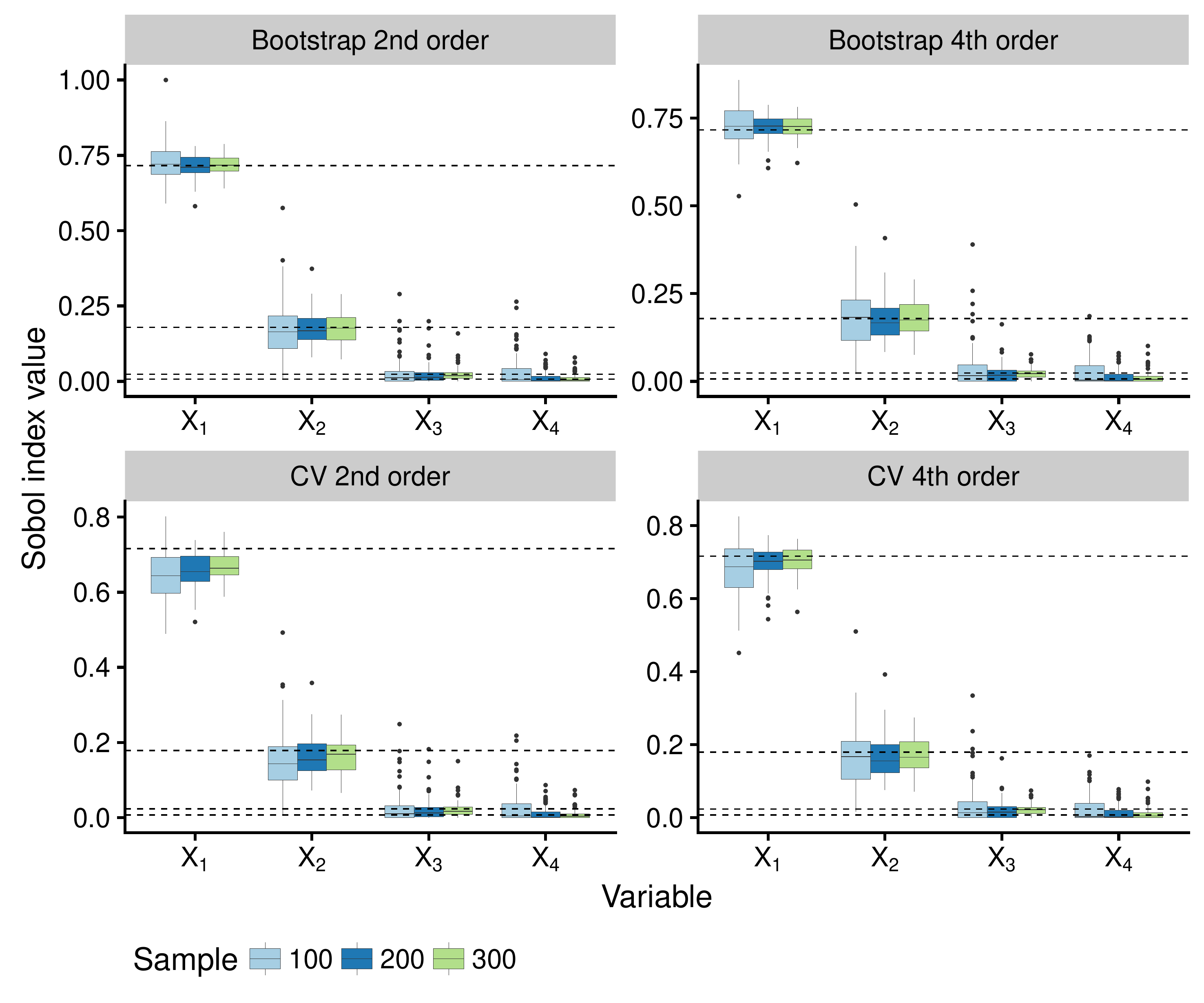}
	\caption{Estimated values from the 100 iterations for the
		\textit{g}-Sobol model using the cross-validation and bootstrap
		procedures to estimate the bandwidth. Both methods were calculated using
		an Epanechnikov kernel of second and fourth order. The horizontal dashed
		lines represent the theoretical values for each Sobol index.}
\end{figure*}

Figure~\ref{fig:all-sobol} and~\ref{fig:np-sobol} presents the estimated
Sobol indices for the \textit{g}-Sobol model. The first Figure presents
the indices using all the algorithms described in the last paragraph.
The next Figure examines further the bandwidth adjust between the
classic cross-validation and the bootstrap methods.

Measuring the bandwidth with the classic cross-va\-li\-da\-tion
procedure, the bias with the second order kernel is greater than with
the fourth order kernel. The behavior is not surprising. In the latter
case, the regression assumes that the inherent curve \(\mathbb{E}[Y\vert
	X]\) has at least four finite derivatives and the bias has a better
adjustments due to the smoothness.

The proposed bootstrap algorithm reduces the bias in all the cases
giving an approximate value near to the real one. It overestimates the
regression curve by selecting a bandwidth smaller. This over-fitting
causes that variance increases and the Sobol index gets larger. Notice
that in most cases the procedure corrects the structural bias. However,
for a fourth order kernel, the bias will be already controlled with the
classic cross-validation procedure. Therefore, the proposed method will
raise the values, causing a Sobol index overestimation.

For all variables, the non-parametric methods achieve the theoretical
values, compared with the other methodologies.

The Table~\ref{tab:sobolComparison} presents the bias and variance  of
$\widehat{S}_{i}^{\mathrm{Boot}}$  and  $\widehat{S}_{i}^{\mathrm{CV}}$.  Notice
how the bias using a second order kernel with the bootstrap method is lower with
respect to the cross-validation counterpart. If we use the fourth order
kernel, the bias of the bootstrap method increases, while for the
cross-validation remains under the true value of the Sobol index $S_i$.  Recall
our procedure over-fits the cross-validation procedure to reduce the bias. One
disvantange of our procedure is a slightly increasing in the variance. In
Figure~\ref{fig:boot-np-mean} we observe how the variance oscillates among each
iteration.

In the bottom of Table~\ref{tab:sobolComparison} we present the average raw
distance between the bootstrap against the cross-validation estimators for
$S_i$. The results show that in average   $\widehat{S}_{i}^{\mathrm{Boot}}$ is
over $\widehat{S}_{i}^{\mathrm{CV}}$ most of the cases.
Figure~\ref{fig:np-sobol} confirms this behavior.

\begin{table}[htpb]
	\centering
	\begin{tabular}{cccccccc}
		\toprule
		&
		& \multicolumn{2}{c}{n = 100}     &
		\multicolumn{2}{c}{n = 200}   &
		\multicolumn{2}{c}{n = 300}  \\
		\cmidrule(r){3-4}
		\cmidrule(r){5-6}
		\cmidrule(r){7-8}
		&

		Variable &
		2$^\mathrm{nd}$ order & 4$^\mathrm{th}$ order & 2$^\mathrm{nd}$ order &
		4$^\mathrm{th}$ order & 2$^\mathrm{nd}$ order & 4$^\mathrm{th}$ order\\
		\midrule
		\midrule
		$\mathrm{Bias}(\widehat{S}_i^\mathrm{Boot})$                             & $X_1$ & 0.0076  & 0.0112  & -0.0021 & 0.0072  & 0.0015  & 0.0088  \\
		                                                                         & $X_2$ & -0.0052 & 0.0044  & -0.0044 & -0.0050 & -0.0026 & 0.0010  \\
		                                                                         & $X_3$ & 0.0057  & 0.0130  & 0.0000  & -0.0015 & -0.0010 & -0.0007 \\
		                                                                         & $X_4$ & 0.0221  & 0.0198  & 0.0047  & 0.0062  & 0.0028  & 0.0040  \\
		\midrule
		$\mathrm{Bias}(\widehat{S}_i^\mathrm{CV})$                               & $X_1$ & -0.0700 & -0.0356 & -0.0581 & -0.0176 & -0.0476 & -0.0120 \\
		                                                                         & $X_2$ & -0.0244 & -0.0092 & -0.0175 & -0.0147 & -0.0133 & -0.0073 \\
		                                                                         & $X_3$ & 0.0024  & 0.0099  & -0.0020 & -0.0027 & -0.0028 & -0.0018 \\
		                                                                         & $X_4$ & 0.0181  & 0.0175  & 0.0036  & 0.0054  & 0.0019  & 0.0035  \\
		\midrule
		\midrule
		$\mathrm{Var}(\widehat{S}_{i}^\mathrm{Boot})$                            & $X_1$ & 0.0039  & 0.0031  & 0.0013  & 0.0011  & 0.0009  & 0.0008  \\
		                                                                         & $X_2$ & 0.0089  & 0.0071  & 0.0030  & 0.0032  & 0.0023  & 0.0025  \\
		                                                                         & $X_3$ & 0.0022  & 0.0036  & 0.0011  & 0.0007  & 0.0004  & 0.0002  \\
		                                                                         & $X_4$ & 0.0025  & 0.0016  & 0.0004  & 0.0004  & 0.0002  & 0.0003  \\
		\midrule
		$\mathrm{Var}(\widehat{S}_{i}^\mathrm{CV})$                              & $X_1$ & 0.0047  & 0.0059  & 0.0019  & 0.0018  & 0.0012  & 0.0012  \\
		                                                                         & $X_2$ & 0.0068  & 0.0065  & 0.0027  & 0.0030  & 0.0021  & 0.0024  \\
		                                                                         & $X_3$ & 0.0017  & 0.0029  & 0.0009  & 0.0006  & 0.0004  & 0.0002  \\
		                                                                         & $X_4$ & 0.0018  & 0.0013  & 0.0003  & 0.0004  & 0.0002  & 0.0003  \\
		\midrule
		\midrule
		$\mathbb{E}(\widehat{S}_{i}^\mathrm{Boot}-\widehat{S}_{i}^\mathrm{CV}) $ & $X_1$ & 0.0777  & 0.0467  & 0.0560  & 0.0248  & 0.0491  & 0.0208  \\
		                                                                         & $X_2$ & 0.0192  & 0.0136  & 0.0131  & 0.0097  & 0.0108  & 0.0082  \\
		                                                                         & $X_3$ & 0.0034  & 0.0031  & 0.0020  & 0.0012  & 0.0018  & 0.0011  \\
		                                                                         & $X_4$ & 0.0040  & 0.0023  & 0.0011  & 0.0008  & 0.0008  & 0.0006  \\
		\bottomrule
	\end{tabular}
	\caption{Mean squared error over the 100 replications  of the estimated against
		the theoretical values for the first four variables of the \emph{g}-Sobol
		model. Here $\widehat{S}_{i}^\mathrm{Boot}$ and $\widehat{S}_{i}^\mathrm{CV}$ are
		the estimated Sobol indices using the Bootstrap and Cross-validation
		methods. The last four rows estimate the average distance between the
		Bootstrap and Cross-validation estimators.}\label{tab:sobolComparison}
\end{table}

Table~\ref{tab:bw} presents the median estimated bandwidths for the
\textit{g}-Sobol. The algorithm calculated the bandwidths using
cross-validation and the bootstrap methods with second order
Epanechnikov kernel. The results show us the over-fitting explained
before, due to the choice of smaller bandwidths for the bootstrap
algorithm. Here, there were values that did not converge to an optimal
solution and the bandwidth \(h\) tend to infinity. The phenomenon is due
to the regression curve \(\mathbb{E}[Y\vert X_{i}]\) is almost flat,
causing that their variance stay in almost zero. For those examples, the
non-parametric curve estimator represent only the mean of the data
regarding \(Y\).

\begin{table}[htp]
	\centering
	\begin{tabular}{ccccc}
		\toprule
		Method           & Bandwidth & $n=100$ & $n=200$ & $n=300$
		\\
		\midrule
		Bootstrap        & \(h_{1}\) & 0.005   & 0.004   &
		0.004                                                                     \\
		                 & \(h_{2}\) & 0.012   & 0.009   & 0.008   \\
		                 & \(h_{3}\) & 0.037   & 0.019   & 0.020   \\
		                 & \(h_{4}\) & 0.130   & 0.066   & 0.032   \\
		\midrule
		Cross-validation & \(h_{1}\) & 0.051   & 0.045   &
		0.039                                                                     \\
		                 & \(h_{2}\) & 0.096   & 0.079   & 0.065   \\
		                 & \(h_{3}\) & 0.187   & 0.121   & 0.129   \\
		                 & \(h_{4}\) & 0.373   & 0.226   & 0.168   \\
		\bottomrule
	\end{tabular}
	\caption{Median bandwidths estimated from the 100 iterations for the
		first four variable of the \textit{g}-Sobol model.}\label{tab:bw}
\end{table}

\subsection{Hydrologic application}%
\label{sec:dyke}
One academic real case model to test the performance in sensitivity
analysis is the dyke model. This model simplifies the 1D hydro-dynamical
equations of Saint Venant under the assumptions of uniform and constant
flow rate and large rectangular sections.

The following equations recreate the variable \(S\) which measures the
maximal annual overflow of the river (in meters) and the variable
\(C_{p}\) which is the associated cost (in millions of euros) of the dyke.
\begin{align}
	\label{eq:dyke-S}
	S     & = Z_{v}+H-H_{d}-C_{b}                              \\
	      & \quad\text{with}\quad H=\left(\frac{Q}{BK_{s}
		\sqrt{\frac{Z_{m}-Z_{v}}{L}}}\right) \nonumber                   \\
	\label{eq:dyke-Cp}
	C_{p} & = \bm{1}_{S>0} + \left[0.2 + 0.8
		\left(1-\exp{\frac{-1000}{S^{4}}}\right)\right] \bm{1}_{S\leq 0} \\
	      & \qquad + \frac{1}{20} \left(H_{d} \bm{1}_{H_{d}>8}
	+ 8 \bm{1}_{H_{d}\leq 8}  \right) \nonumber
\end{align}

Table~\ref{tab:inputs-dyke} shows the inputs (\(p=8\)). Here
\(\bm{1}_{A}(x)\) is equal to 1 for \(x\in A\) and 0 otherwise. The variable
\(H_{d}\) in Equation~\eqref{eq:dyke-S} is a design parameter for the
Dyke's height set as a \(\mathrm{Uniform}(7,9)\).

In Equation~\eqref{eq:dyke-Cp}, the first term is 1 million euros due to
a flooding (\(S>0\)), the second term corresponds to the cost of the dyke
maintenance (\(S\leq 0\)) and the third term is the construction cost
related to the dyke. The latter cost is constant for a height of dyke
less than 8 m and is growing like the dyke height otherwise.
\begin{table*}
	\centering
	\begin{tabular}{clcl}
		\toprule
		Input   & Description                 & Unit          & Probability
		Distribution                                                                    \\
		\midrule
		\(Q\)   & Maximal annual flowrate     & m\({}^{3}\)/s & \(\mathrm{Gumbel}(1013,
		558)\) truncated on [500, 3000]                                                 \\
		\(K_s\) & Strickler coefficient       & ---           & \(\mathcal{N}(30, 8)\)
		truncated on \(\left[15, \infty\right)\)                                                   \\
		\(Z_v\) & River downstream level      & m             &
		\(\mathrm{Triangular}(49, 50, 51)\)                                             \\
		\(Z_m\) & River upstream level        & m             &
		\(\mathrm{Triangular}(54, 55, 56)\)                                             \\
		\(H_d\) & Dyke height                 & m             &
		\(\mathrm{Uniform}(7,9)\)                                                       \\
		\(C_b\) & Bank level                  & m             &
		\(\mathrm{Triangular}(55, 55.5, 56)\)                                           \\
		\(L\)   & Length of the river stretch & m             &
		\(\mathrm{Triangular}(4990, 5000, 5010)\)                                       \\
		\(B\)   & River width                 & m             &
		\(\mathrm{Triangular}(295, 300, 305)\)                                          \\
		\bottomrule
	\end{tabular}
	\caption{Input variables and their probability distributions.}%
	\label{tab:inputs-dyke}
\end{table*}

For a complete discussion about the model, their parameters and their
meaning the reader can review \citet{Iooss2015}, \citet{Rocquigny2006}
and their references.

\begin{figure*}[htp]
	\centering
	\centering \includegraphics[width = \textwidth]{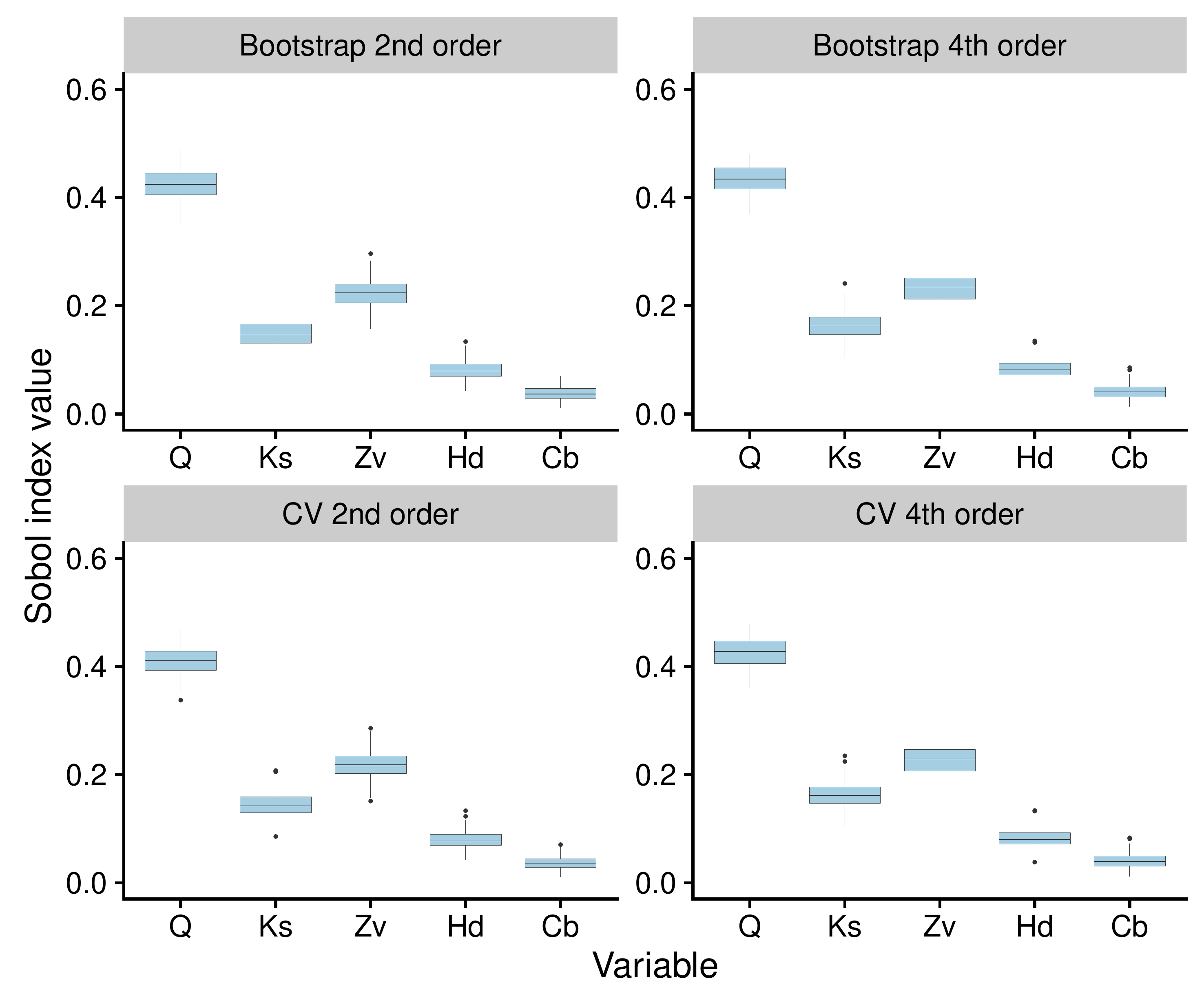}
	\includegraphics[width = \textwidth]{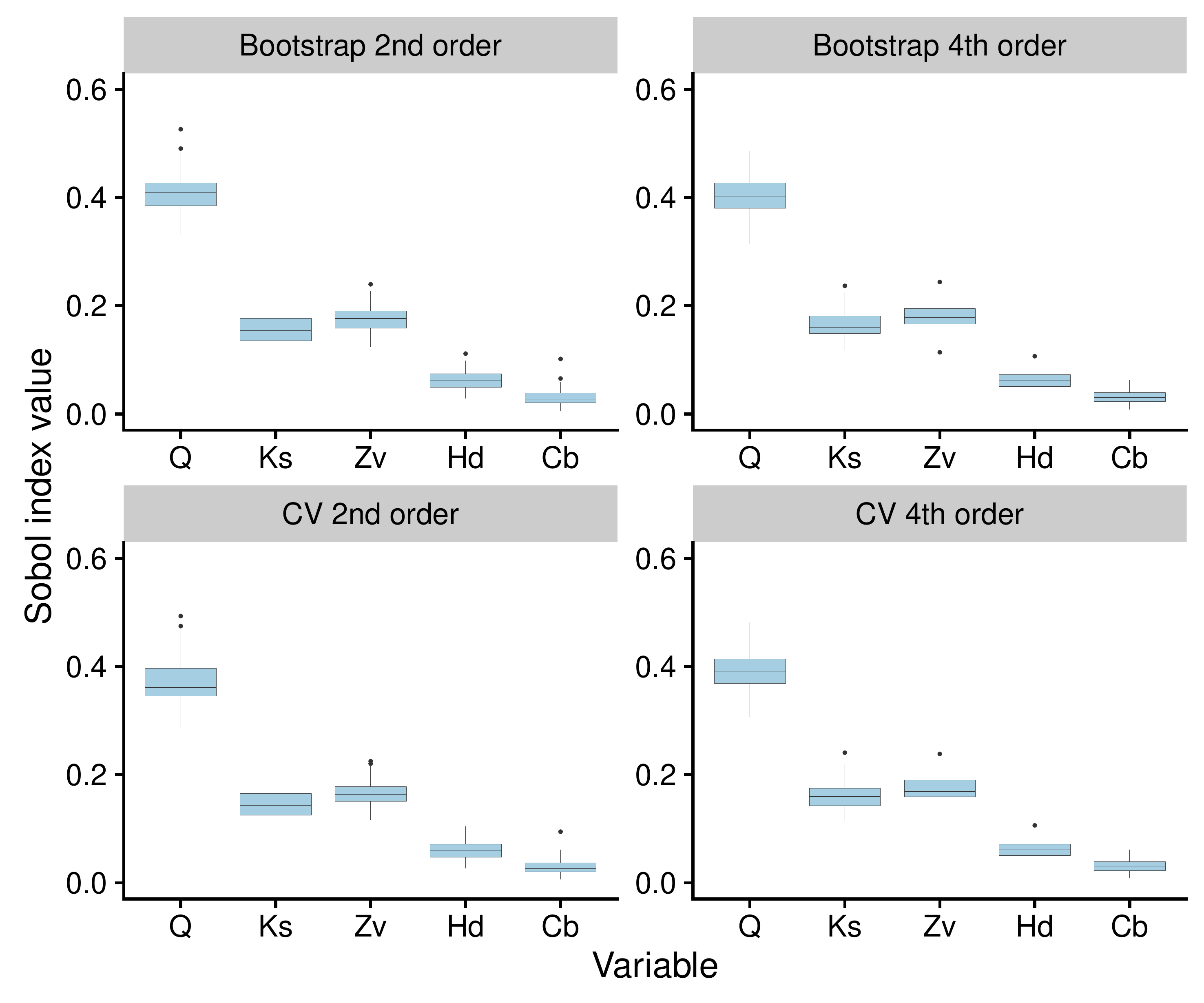}
	\caption{Estimated values from the 100 iterations for the output \(S\) and
	\(C_{p}\) in the Dyke model.}%
	\label{fig:all-dykeS}
\end{figure*}


We generated 1000 observations for each input according to
Table~\ref{tab:inputs-dyke} and their respective values for \(S\) and
\(C_{p}\). Figure~\ref{fig:all-dykeS} shows the result of simulations for
the output \(S\) and \(C_{p}\) of the Dyke model using the cross-validation
and bootstrap procedures.

For both output \(S\) and \(C_{p}\) we see that the variables in order of
importance are \(Q\), \(Z_{v}\), \(K_{s}\), \(H_{d}\) and \(C_{b}\). The rest of
variables have values near to zero, and they provided insignificant
impact to the output.

As reported in Table~\ref{tab:comparison}, we compared the values from
our procedure against the reported by \citeauthor{Iooss2015}. The values
of the Sobol indices detect the influence of each variable compared with
the classic Monte-Carlo and meta-models procedures. The exception is
\(H_{d}\), which in our case decreased to values near to \(5\%\) against the
reported values of \(12.5\% - 13.9\%\).
\begin{table}
	\centering
	\begin{tabular}{lccccc}
		\toprule
		Indices (in \%)               & \(Q\) & \(K_{s}\) & \(Z_{v}\) & \(H_{d}\) &
		\(C_{b}\)                                                                       \\
		\midrule
		\(S_{i}\) Monte-Carlo (Iooss) & 35.5  & 15.9      & 18.3      & 12.5      & 3.8
		\\
		\(S_{i}\) Meta-model (Iooss)  & 38.9  & 16.8      & 18.8      & 13.9      & 3.7
		\\
		\midrule
		\(S_{i}\) Bootstrap           & 40.5  & 15.5      & 18.1      & 5.7       & 2.9
		\\
		\(S_{i}\) Cross-validation    & 37.2  & 14.6      & 17.2      & 5.5       & 2.8
		\\
		\bottomrule
	\end{tabular}
	\caption{Comparison between the Sobol indices in the dyke model reported
	by \citet{Iooss2015} and our method. The Monte-Carlo and meta-model
	methods used samples of \(10^{5}\). The bootstrap and cross-validation
	method used samples of \(10^{3}\). In all cases the simulation repeated
	the experiment 100 times.}\label{tab:comparison}
\end{table}

\section{Conclusions}\label{sec:conclusions}
This paper presented an alternative way to estimate first order Sobol
indices for the general model \(Y = \varphi(X_{1}, \ldots, X_{p})\). These
indices are calculated using the formula \(S_{i} = \mathrm{Var}(\mathbb{E}[Y\vert
	X_{i}])/\mathrm{Var}(Y)\). The method builds the regression curve
\(\mathbb{E}[Y\vert X_{i}]\) by a kernel non-parametric regression.

The least-square cross-validation procedure is a classic way to find the
bandwidth. However, the literature presents cases where there exist a
finite-sample bias on the model. One way to correct is increasing the
number of samples. This method proposes a bootstrap algorithm to correct
the bias, by first estimating the normalized residuals of the model and
then recreating a bootstrap version of the response variable. With this
new data, the algorithm estimates an empirical version of the least
squared error. We call it \textit{Bootstrap Least-Square} criterion and
denoted \(\mathrm{BLS}(h)\). The function \(\mathrm{BLS}(h)\) finds its
minimum in a value \(\hat{h}_{boot}\).

The proposed algorithm over fits the regression curve \(\mathbb
{E}[Y\vert X_{i}]\), because it chooses a smaller bandwidth to increase
the variability of the curve. It approximates the first order
Sobol indices, but it could overestimate them when using fourth order
kernels. The method proposed reduces the structural bias of caused for the non-parametric estimator. However, due to its construction the estimators have a slightly increased variance. 

The function \(\mathrm{BLS}\) was minimized using a Brent-type routine,
implemented in the \textit{R} function \texttt{optimize}. Due to the
complexity of the target function, one future improvement to the
algorithm is to use a global minimizer like simulated annealing to
compare the results. In this scenario, we will expect a better choice of
the bandwidths and observe a better adjust for the Sobol indices.

The method showed a consistent approximation to the Sobol indices only
having the observational data available. In all cases, the
non-parametric estimator using cross-validation and bootstrap
approximate the influential variables in the \(g\)-Sobol and Dyke models.

We consider only the indices with simple interactions between one
variable with respect the output. The hig\-her order indices and total
effects will remain for a further study. We will estimate the
multivariate non-parametric surface for multiple variables. Then, we
have to approximate the surface variability over some range. The latter
step will be an interesting topic of study due to the numerical
complexities.

\section*{Funding details}\label{sec:ack}
I acknowledge the financial support from the Vicerrectoría de Investigación
de la Universidad de Costa Rica through the project 821--B4--257 administrated
by the Escuela de Matemática and Centro de Investigaciones en Matemática Pura y
Aplicada (CIMPA).

\section*{Disclosure statement}

No potential conflict of interest was reported by the author.

\bibliographystyle{apalike} %
\bibliography{sobol.bib}

\end{document}